\documentstyle[prb,aps,multicol,epsf]{revtex}
\begin{document}

\author{Y. De Wilde, H. Enriquez, N. Bontemps} \address{Laboratoire de
Physique de la Mati\`{e}re Condens\'{e}e, Ecole Normale Sup\'{e}rieure, 24 rue Lhomond,
75231 Paris Cedex 05, France}
\author{T. Tamegai} \address{Department of Applied Physics, The Univeristy of Tokyo, Hongo,
Bunkyo-ku, Tokyo 113-8656, Japan} 
\title{Microwave Induced Instability Observed in BSCCO 2212 in a Static
Magnetic Field}

\maketitle

\begin{abstract}
We have measured the microwave dissipation at 10 GHz through the imaginary part of the 
susceptibility,  $\chi^"$, in a BSCCO 2212 single crystal in an external static magnetic field $H$ 
parallel to the c-axis at various fixed temperatures. The  characteristics of $\chi^"(H)$ exhibit a sharp 
step at a field $H_{step}$ which strongly depends on the amplitude of the microwave excitation 
$h_{ac}$. The characteristics of  $h_{ac}$ vs. $H_{step}$, qualitatively reveal the behavior expected 
for the magnetic field dependence of Josephson coupling.

\end{abstract}

%\pacs{PACS numbers: 74.25.Nf, 74.72.Bk, 74.50.+r}

\begin{multicols}{2}
Vortex physics on high critical temperature superconductors (HTSC) has driven a lot of interest over 
the past few years\cite{Crabtree97}. Because of  the short coherence length which characterizes these 
systems,  some HTSC such as BSCCO 2212, have to be described as stacks of intrinsically Josephson 
coupled superconducting planes\cite{Kleiner92}, rather than as superconductors with a high effective 
mass anisotropy. In an  externally applied magnetic field, DC 
transport measurements \cite{DCmeasurements} and thermodynamic \cite{thermodynamic} studies  
have shown that melting  of vortices is a first order process. In Josephson coupled HTSC's , it is still 
not clear if melting goes along with a total decoupling of the superconducting planes, or if 
instead decoupling is only achieved beyond the melting transition. Experimental  studies of the 
Josephson plasma resonance (JPR) have been performed to investigate the temperature and magnetic 
field  dependence of  Josephson coupling\cite{Tsui94,Matsuda97,Shibauchi98,Shibauchi99}. The 
results point to a drop of Josephson coupling at the vortex  melting \cite{Shibauchi98,Shibauchi99}. 
Other JPR experiments on BSCCO 2212 have shown the possible occurrence of nonlinear effects 
driven by a microwave field \cite{Hanaguri98}.

We have recently presented a detailed study of melting of  vortices in a BSCCO 2212 single crystal 
via microwave dissipation measurements\cite{Enriquez98} at 10 GHz. While a careful study of vortex 
melting requires current densities as small as possible, and hence low microwave excitations,  in this 
paper we focus on what happens when the amplitude of the microwave field $h_{ac}$ is progressively 
increased. We observe a novel feature which shows up as a sharp step in the 
characteristics of dissipation vs. magnetic field at fixed temperature. In contrast with vortex melting, 
whose field position, $H_m$, only depends on the temperature, the field value at which the step 
occurs, $H_{step}$, drastically depends on  $h_{ac}$,  which we argue is closely related to the 
Josephson interlayer phase coherence $<cos\phi>$. 

 The experimental set-up that we use to measure the 10 GHz microwave dissipation has been 
described 
elsewhere \cite{Enriquez96}. The sample is placed in a 
TE$_{102}$ resonant cavity (resonance frequency $\omega / 2 \pi = 9.6$ GHz) of an Electron Spin 
Resonance (ESR) ESP 300 E Br\"{u}ker spectrometer, at the maximum of the microwave magnetic 
field $h_{ac}(\omega)=h_{ac}\cos \omega t$ and in zero electric field.  The external static magnetic 
field $ H$ is applied parallel to the c-axis of the crystal, while $h_{ac}(\omega)$ is oriented 
within the ab-plane. The microwave field $h_{ac}(\omega)$ induces surface currents in the sample, 
which in this geometry flow within the ab-plane and along  the c-axis. The surface 
currents are responsible for an energy dissipation which is measured through the 
imaginary part of the susceptibility $\chi^"$. With a small $h_{ac}$, in the range of a few mOe, 
melting of the vortex solid in BSCCO 2212 shows up as an increase of dissipation in the 
characteristics of $\chi^"$ vs. $H$ recorded at a fixed temperature \cite{Enriquez98}. 

 The BSCCO 2212 single crystal used in the experiments is a rectangular platelet with dimensions $2 
\times 0.8 \times 0.03$ mm$^3$. 
For this sample, the critical temperature $T_c=84$ K with a transition width $\Delta T_c \approx 
0.4$ K, as measured at 10 GHz. An attenuator placed between the source and the cavity enables a 
step-by-step variation of  the microwave power in the cavity, with an attenuation of  1dB between each 
step. An attenuation of 1 dB corresponds to a reduction of $h_{ac}$ by a factor 1.12. Fig. 
\ref{Jump65K} shows a typical $\chi^"$ vs. $H$ characteristic obtained at $T=65$K with  an 
input power of 0.8 mW, corresponding to $h_{ac}\approx 60$ mOe. Here melting occurs 
around $H_m=315$ Oe and shows up as a small jump and a change of slope in $\chi^"(H)$. There is 
no influence of the input power on the value of $H_m$ measured at 10 GHz with 
our technique. Good agreement is found between the $H_m(T)$ phase diagram measured at 10 
GHz and that obtained from DC SQUID measurements on the same sample \cite{Enriquez98}. The 
small jump in  $\chi^"$ at $H_m$  is analogous to the sharp onset of resistivity observed in DC 
transport measurements \cite{DCmeasurements}. A salient new feature in Fig.\ref{Jump65K}, is the 
sharp step which corresponds to a sudden increase of dissipation at a field $H_{step}\approx 200$ Oe. 
The curves shown in the inset of  Fig. \ref{Jump65K} were 
obtained at a temperature of 65K, with $h_{ac}\approx 47$mOe, when sweeping the magnetic field 
up and down successively. A hysteretic behavior of the step is observed. The amplitude of the 
hysteresis loop at 65 K is of the order  of 10 Oe. 

Fig. \ref{PowerDep} shows a series of characteristics of $\chi^"$ vs. $H$ recorded at 65 K, where the 
input power has systematically been attenuated by 2 dB between subsequent recordings in order to 
vary $h_{ac}$. In contrast with $H_m$, the field $H_{step}$ strongly depends on the input power. At  
20 dB, which corresponds to $h_{ac}\approx 95$mOe, the step occurs in the solid phase, $H_{step}$ 
well below $H_m\approx 315$ Oe, and shifts to higher fields as the attenuation is increased, i.e. as 
$h_{ac}$ is reduced.  The step survives in the liquid phase beyond $H_m$, where it presents  
significant broadening. Similar observations of a step 
with 10 GHz dissipative measurements were done on other BSCCO 2212 samples, showing the 
intrinsic origin of the feature. On a sample with a somewhat broader 
superconducting transition, the step was broader and reduced in amplitude compared to the results 
presented in Fig. \ref{Jump65K} and Fig. \ref{PowerDep}. The observation of the step seems to 
require samples with very homogeneous superconducting properties. At high enough input power, the 
characteristics of  $\chi^"$ vs. $H$ exhibit an offset, indicating that the extra dissipation produced by 
the step {\it has already occurred} at $H=0$. Other measurements of  high purity BSCCO 2212 single 
crystals placed in a 10 GHz niobium resonator 
\cite{Jacobs96} in zero field showed switching of the surface resistance, probably of  the 
same origin as the step in $\chi^"(H)$  reported here.  

In our geometry, $h_{ac}(\omega)$ tend to tilt slightly the field on the sample away from the c-axis of 
the crystal. In tilted magnetic fields,   vortices in samples with a large anisotropy such as BSCCO 
2212 adopt a tilted structure made of  vortex pancakes linked by Josephson vortex segments  
\cite{Bulaevskii92,Feinberg}, or may even form a two components structure with coexistent 
Abrikosov and Josephson vortex systems, orthogonal to each other \cite{Benkraouda95,Koshelev99}. 
While the density of pancake vortices depends on the static field component $H$, the density of 
Josephson strings, regardless of  the details of the vortex structure,  is driven by the 
field component parallel to the superconducting CuO$_2$ planes, here 
$h_{ac}(\omega)$. The step amplitude in $\chi^"(H)$ is proportional to the microwave input power 
and independent of the static field component. This points to a  dissipation process related to 
Josephson strings induced by the microwave field and  not to  pancake vortices. 

We have thoroughly investigated the  temperature and power dependence of the step between 75 K 
and 30 K, changing the temperature by steps of  5K between each series of recordings, and the 
attenuation of the microwave excitation by steps of 1dB.  Fig. \ref{Kink_h_H_melt}  shows two series 
of plots of  
$h_{ac}$ vs.  $H_{step}$, which cover the entire temperature range of our studies. The arrows in 
these plots indicate the melting transition at $H_m$, whenever observed.  Fig. \ref{Kink_h_H_melt}  
shows that $h_{ac}$ is a decreasing function of  the static field . Down to about 45 K,  the 
characteristics of  $h_{ac}$ vs. $H_{step}$ exhibit a fairly distinct drop which in each case coincides 
with the melting transition.  The dependence of  $h_{ac}$ vs. magnetic field is qualitatively the 
same as that expected for the interlayer Josephson coupling. The additional phase difference 
$\phi$ due to the presence of pancake vortices in a  perpendicular static magnetic field $H$ produces 
a loss of  interlayer phase coherence  $<cos\phi>$, and so reduces  the average Josephson coupling 
energy (or equivalently, the average Josephson current) when $H$ increases \cite{Koshelev96}. 
Extra decoupling  of the layers seems to occur  at melting, producing a sudden drop of  $<cos\phi>$. 
Experimental observation of  this drop has been reported in the organic superconductors $\kappa$-
(BEDT-TTF)$_2$Cu[N(CN)$_2$]Br and, recently, in BSCCO 2212 in measurements of  the 
magnetic field dependence of the JPR frequency \cite{Shibauchi98,Shibauchi99}$\omega_{pl}$. Note 
that at $T=40$K and below, melting could not be resolved in our measurements, neither in the 
characteristics of $\chi^"$ vs. $H$, nor via magnetization measurements with a SQUID 
magnetometer. 
Below 40 K, the drop in the characteristics of  $h_{ac}$ vs. $H_{step}$ does not seem to be present 
as well in Fig. \ref{Kink_h_H_melt}, in contrast with recent JPR data in which a drop in 
$\omega_{pl}^2$ vs. $H$ was still observed below the critical point \cite{Shibauchi99}. 

Fig. \ref{Kink_h_H_melt} shows that in the liquid phase $h_{ac}$ can be fitted with a  
$1/H^\alpha$ dependence. Down to 50 K, the fits give $\alpha\approx 0.8$. Approaching 40 K, and 
below, $\alpha$ progressively increases and is of order 1.2 around 30 K.  Koshelev\cite{Koshelev96} 
has predicted that in the particular case of  a vortex liquid, $<cos\phi>$ should exhibit a $1/H^\alpha$ 
dependence, with $\alpha$ close to unity, and be proportional to the square of  the JPR frequency 
$\omega_{pl}^2$ ; these predictions have been confirmed in JPR 
studies\cite{Matsuda97,Shibauchi98,Shibauchi99}. An 
increase of $\alpha$ is expected for $<cos\phi>$ as a consequence of disorder in the pancake 
arrangement along the   c-axis \cite{Bulaevskii96}, due for instance to an increase of pinning at low 
temperature. Fig. \ref{Kink_h_H_melt}  shows that the relation  between $h_{ac}$ and $H$ found in 
BSCCO is very  close to that expected for $<cos\phi>$ in the liquid phase. It suggests that  $h_{ac}$ 
provides a direct measure of  the interlayer phase coherence,  being close to proportional to 
$<cos\phi>$, in a similar way as $\omega_{pl}^2$. This is supported by the close similarity existing 
between  the normalized characteristics of   $h_{ac}$ versus $H_{step}$, shown in Fig. 
\ref{Kink_h_H_melt}c, and the JPR data presented in ref. \cite{Shibauchi99} . Within this scenario, 
the  fact that $h_{ac}$ still has a substantial non zero value right beyond  $H_m$ in Fig. 
\ref{Kink_h_H_melt} indicates that phase coherence is not completely lost when the vortices enter the 
liquid phase. It is only at higher  fields that vanishingly small $h_{ac}$'s probe almost totally 
decoupled layers. In the characteristics of   $\chi^"$ vs. $H$ shown in Fig. \ref{PowerDep}, the step 
progressively broadens in the liquid phase, suggesting  that the phase difference between the 
superconducting layers becomes less and less homogeneous over the sample thickness as the static 
field is increased beyond $H_m$. 

The step observed in Fig. \ref{Jump65K} and Fig. \ref{PowerDep} indicates that a new dissipation 
process sets in at once  when  $H$ reaches a threshold value $H_{step}(h_{ac})$. 
When a large area Josephson junction is subject to a small in-plane magnetic field $h$, penetration of  
Josephson strings is prevented if the supercurrent  density circulating at the junction periphery, over a 
thickness given by the London penetration depth $\lambda_J$, is smaller than the Josephson critical 
current density  \cite{Tinkham96} $j_c$. This defines the  lower Josephson critical field  $h_{c1{\bf  
J}}\propto j_c  \lambda_J $ below which $h$ is screened. Similarly, a stacked structure of Josephson 
junctions such as a BSCCO single crystal should expel  in-plane microwave fields of amplitude 
$h_{ac}$ smaller than $h_{c1{\bf J}}$ over a length scale $\lambda_J $ (generally called $\lambda_c 
$ ). An additional static  field component $H$ perpendicular to the Josephson stacks produces 
pancake vortices in the superconducting planes, which result in an extra loss of interlayer phase 
coherence and  hence in a decrease of  $h_{c1{\bf  J}}(H)$. In each characteristic of $\chi^"$ vs. $H$  
shown in Fig.\ref{Jump65K} and Fig.\ref{PowerDep}, $h_{ac}$ is kept at a constant value during the 
sweeps of  $H$. Sweeping up the static field $H$, results in a decay of  $h_{c1{\bf  J}}(H)$. When 
$H$ is such that  $ h_{c1{\bf  J}}(H)> h_{ac} $, no microwave induced Josephson strings penetrate 
the sample. A further increase of  $H$ leads to the reversed situation,  $ h_{c1{\bf  J}}(H)<h_{ac} $ ; 
Josephson strings enter and leave then the sample at the microwave frequency, giving rise to an extra 
contribution to the dissipation. This process is expected to occur at a threshold field such that  
$h_{c1{\bf  J}}(H) = h_{ac}$, producing a step in the characteristics of   $\chi^"$ vs. $H$ like that 
observed experimentally at $H=H_{step}$. Hence, a possible explanation for the step could be that it 
is associated with the lower Josephson critical field  $h_{c1{\bf  J}}$ in BSCCO below which 
penetration of Josephson strings is prevented at 10 GHz.

An important consequence of this scenario is that the diagram of $h_{ac}$ vs. $H_{step}$ should 
permit to explore how Josephson coupling evolves in a static perpendicular field $H$, provided that 
the relation between $h_{ac}$ and $<cos\phi>$ is known. As a first guess, we  
take  $j_c \propto h_{ac}/\lambda_J$ , which is justified at the step, since we suppose  
$h_{c1{\bf  J}}(H_{step}) = h_{ac}$ in each characteristic of $\chi^"$ vs. $H$. Like in ordinary 
Josephson junctions, we also  assume \cite{Tinkham96} that $\lambda_J\propto{j_c}^{-\frac{1}{2}} 
\propto<cos\phi>^{-\frac{1}{2}}$. Combined with the $1/H$ dependence of  
$<cos\phi>$ in the decoupled liquid phase predicted by Koshelev \cite{Koshelev96}, one should then 
expect a $1/H^{1/2}$ dependence for $h_{ac}$.The experimental data in Fig. \ref{Kink_h_H_melt}  
show instead a $1/H^\alpha$ behavior with $\alpha$ close to 1. On the other  hand, the losses in the 
liquid phase should  be proportional to the surface resistance $\propto {\rho_{fl}}/{\lambda_J}$, 
where $\rho_{fl}$ is the  flux flow resistivity which varies linearly \cite{Sonin}with $H$. If 
$\lambda_J\propto H^{\frac{1}{2}}$, the losses in the liquid phase should 
then exhibit a $H^{1/2}$ dependence \cite{Sonin} ; instead, the results presented in ref. 
\cite{Enriquez98} show a behavior closer to linear in $H$. These discrepancies point to some 
difficulty essentially related to the field dependence of  $\lambda_J$. The  main reason might be the 
lack of knowledge about either the effective thickness $\lambda_{J}^{eff}$ of the 
layer through which the microwave surface currents flow, or of its magnetic field dependence. The 
microwave loss data from ref.\cite{Enriquez98} suggest that $\lambda_{J}^{eff}$ weakly 
depends \cite{Sonin} on $H$, which would then be consistent with the experimentally observed  
$1/H$ variation of $h_{ac}$. More realistic models, taking into account the influence of effects such 
as surface defects on  $\lambda_{J}^{eff}$ might be able to reconcile the experimental observations 
with theory.  

Recent JPR experiments of Hanaguri {\it et al.}on BSCCO 2212 have shown nonlinear effects 
induced by microwave power \cite{Hanaguri98}. The nonlinearities of the JPR were qualitatively 
explained with the perturbed sine-Gordon equation for the Josephson phase. It predicts a lowering of 
the resonance frequency and the appearance of a hysteretic bistable regime at high microwave current, 
consistent with the experimental JPR data \cite{Hanaguri98,Koshelev_tbp}. Fig.\ref{Kink_h_H_melt} 
shows that, at a fixed value of the static magnetic field, the step occurs at lower temperature when 
$h_{ac}$, and the associated microwave current, is increased.  This behavior is similar to that 
observed 
in ref. \cite{Hanaguri98} for the nonlinear JPR. Hence,  another possible interpretation for the step is 
that it might correspond to the switching between two stable solutions of the perturbed sine-Gordon 
equation. A theoretical estimate of the range of existence of each solution of the sine-Gordon equation 
predicts  \cite{Koshelev_tbp}  $h_{ac}\propto  <cos\phi>^{1/2}$. This is the same analytical 
dependence as that predicted in the scenario involving  the lower critical field $h_{c1{\bf  J}}$. It 
should result in a $1/H^{1/2}$ dependence of  $h_{ac}$ in the liquid phase, 
whereas the experimental data exhibit a behavior closer to $1/H$.  A proper comparison of  our data 
with the nonlinear JPR data \cite{Hanaguri98} and theory requires precise knowledge of the 
longitudinal currents which flow through the superconducting layers in both experiments. These data 
are not available at this stage. 

In summary, the characteristics of $h_{ac}$ vs. $H_{step}$, recorded at various fixed temperatures, 
qualitatively reveal the behavior expected for the magnetic field dependence of Josephson 
coupling ; the characteristics exhibit a drop at $H_m$, which points to  substantial decoupling at the 
melting transition. We propose that the step is either related to the lower Josephson critical 
field  $h_{c1{\bf  J}}$ in BSCCO, or  to nonlinear Josephson plasma resonance. 

We are grateful to A. E. Koshelev for stimulating discussions. This work is supported by CREST and 
Grant-in-Aid for Scientific Research from the Ministry of Education, Science, Sports and Culture of 
Japan.

\begin{minipage}{8.5cm}
\begin{figure}
\epsfxsize=7.5cm
%\epsffile{Figure1.eps}
\caption{Dissipation, $\chi^"$, as a function of the applied magnetic field, $H$, recorded at 65 K with 
a microwave excitation $h_{ac}\approx 60$ mOe. The arrows indicate  the step, at $H=H_{step}$, 
and the melting, at 
$H=H_{melt}$. Inset :  Curves of $\chi^"$ vs.  $H$ resulting from subsequent sweeps up and 
down of the magnetic field $H$ at 65 K with $h_{ac}\approx 47$mOe showing the hysteretic 
behavior 
of  the step.}
\label{Jump65K}
\end{figure}
\end{minipage}
\begin{minipage}{8.5cm}
\begin{figure}
\epsfxsize=7.5cm
%\epsffile{Figure2.eps}
\caption{Series of characteristics of the dissipation $\chi^"$ versus magnetic field $H$ recorded at 65 
K with various input power. The attenuation ranges from 20 dB ($h_{ac}=95$ mOe)  to 40 dB 
($h_{ac}=9.5$ mOe).}
\label{PowerDep}
\end{figure}
\end{minipage}
\begin{minipage}{8.5cm}
\begin{figure}
\epsfxsize=7.5cm
%\epsffile{Figure3.eps}
\caption{Double logarithmic plot of the amplitude of the input excitation, $h_{ac}$, versus the field 
at which the step in dissipation occurs, $H_{step}$, recorded (a) between 75 K and 50 K, and (b)  
between 50 K and 30 K. The arrows locate the melting transition, whenever observed (see text). (c) 
Data corresponding to $T \geq 45$ K plotted on a normalized scale, with 
$h_{ac}$  and $H_{step}$ divided by their values at melting. The dashed line corresponds to 
$h_{ac}\propto 1/ H_{step}$.}
\label{Kink_h_H_melt}
\end{figure}
\end{minipage}

\bibliographystyle{prsty}

\end{multicols}
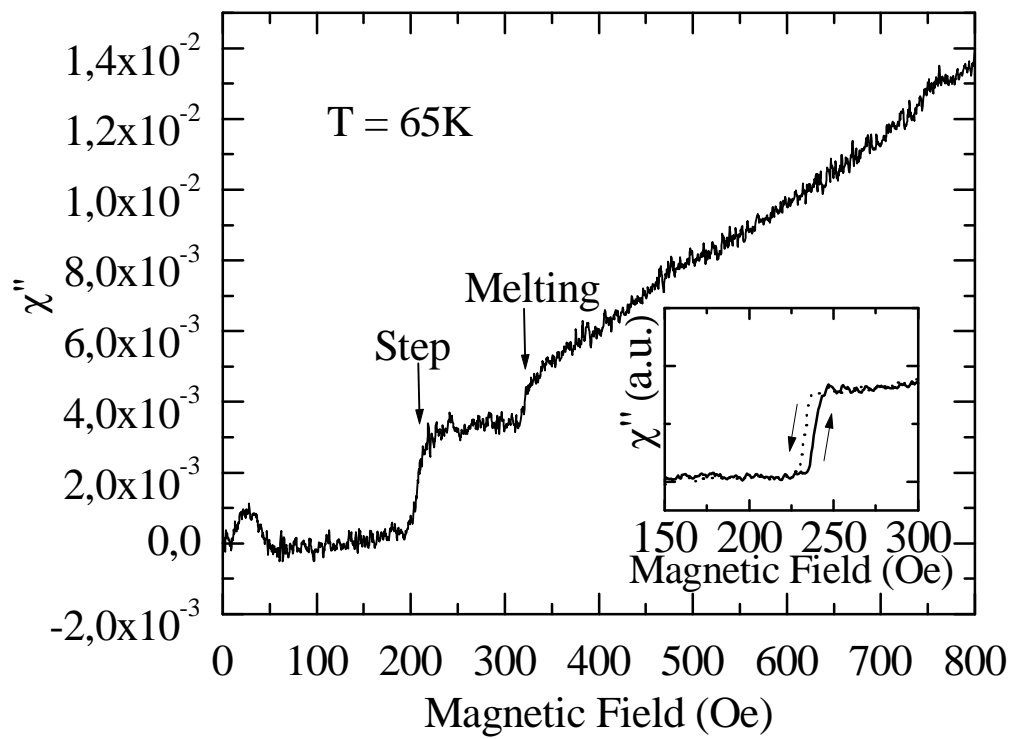
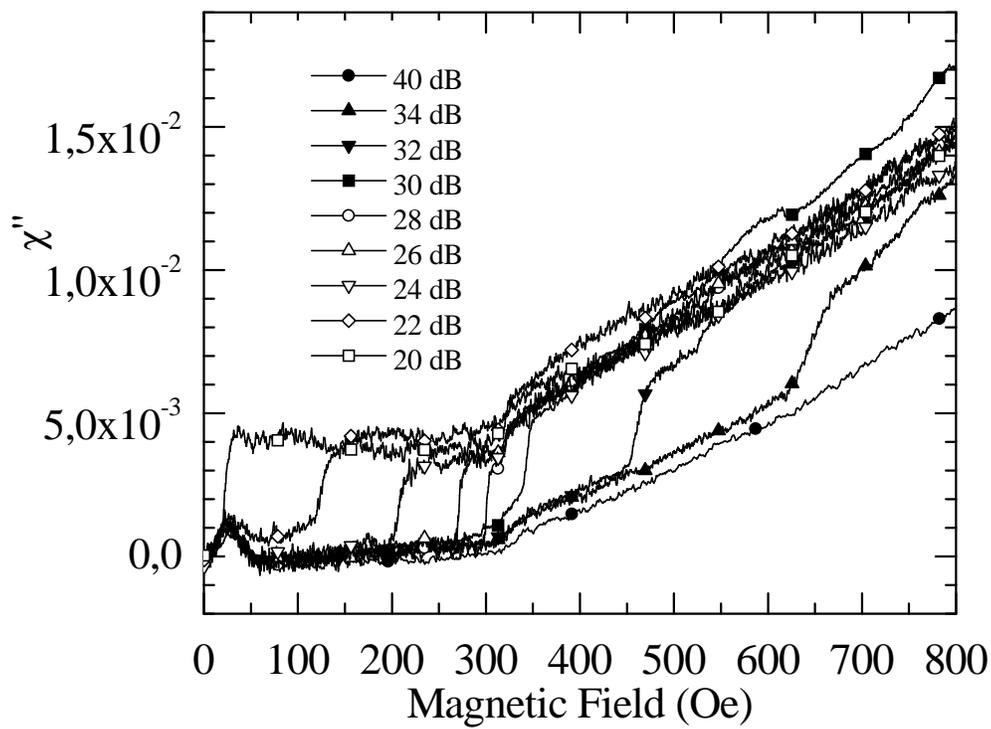
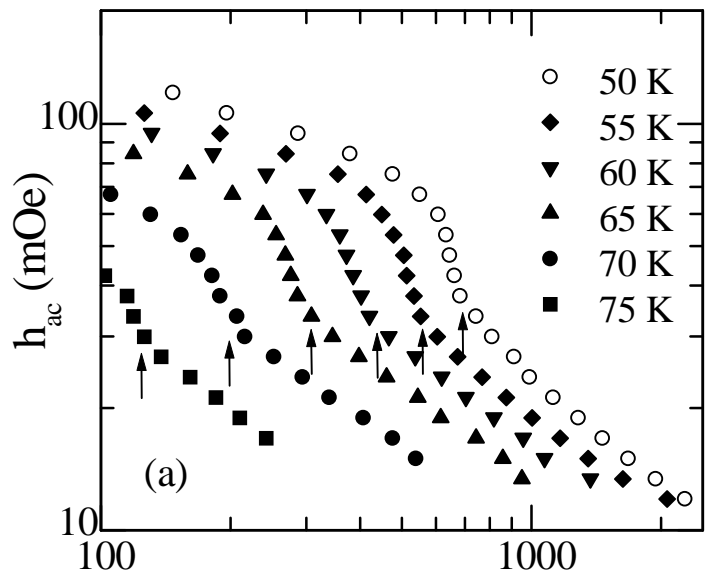
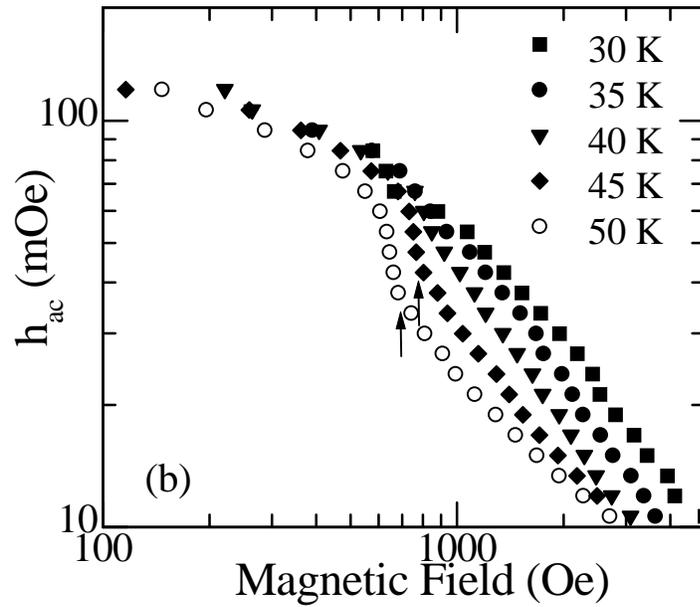
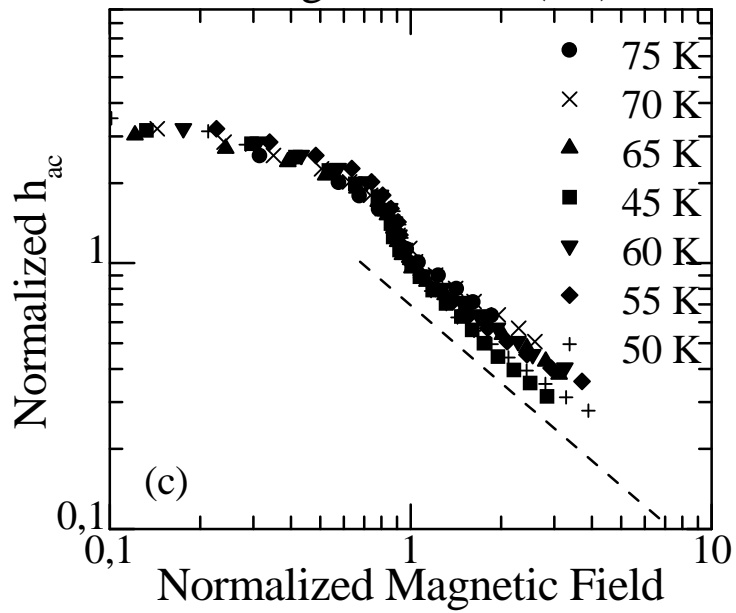
\end{document}